\documentstyle[aas2pp4]{article}
\begin{document}
\pagestyle{plain}

\begin{titlepage}

\title{Diamagnetic blob interaction model of TTauri variability}
\author{Yigal Ultchin{\altaffilmark{1}}, Oded Regev {\altaffilmark{1,2}}}
\and
\author{Claude Bertout{\altaffilmark{2}}}
\altaffiltext{1}{Department of Physics
Technion - Israel Institute of Technology
32000 Haifa, Israel}
\altaffiltext{2}{Laboratoire d'Astrophysique, Observatoire de Grenoble \\
Universit\'e Joseph Fourier,   B.P.53X
38041 Grenoble Cedex, France}

\pagestyle{empty}

\begin{abstract}
{Assuming a diamagnetic interaction between a stellar-spot originated
localized magnetic field and gas blobs in the accretion disk around a T-
Tauri star, we show the possibility of ejection of such blobs out of the
disk plane. Choosing the interaction radius and the magnetic field
parameters in a suitable way gives rise to closed orbits for the ejected
blobs. A stream of matter composed of such blobs, ejected on one side of the
disk and impacting on the other, can form a hot spot at a fixed position on
the disk (in the frame rotating with the star). Such a hot spot, spread
somewhat by disk shear before cooling, may be responsible in some cases for
the lightcurve variations observed in various T-Tauri stars over the years.
An eclipse-based mechanism due to stellar obscuration of the spot is
proposed. Assuming high disk inclination angles it is able to explain many
of the puzzling properties of these variations. By varying the field
parameters and blob initial conditions we obtain variations in the apparent
angular velocity of the hot spot, producing a constantly changing period or
intermittent periodicity disappearance in the models. }
\end{abstract}


\end{titlepage}

\setcounter{page}{1}
\setcounter{footnote}{0}

\section{Introduction}
Most T-Tauri stars (TTS) exhibit irregularities in their luminosity,
observable photometrically and spectroscopically. This variability is
observed on a variety of timescales and in a wide frequency band. Although
the short-term variability can be explained as variable chromospheric
eruption/emission episodes (Worden {\it et al}. 1981), the long-term
variability is not yet fully explained. An extreme example is BP Tau, whose
lightcurve, monitored over 50 days (Simon {et al}., 1990), shows periodic
brightness fluctuations with a period of 7.6d, then an almost constant
brightness level for 2 weeks, and finally the periodic behavior resumes -
but with a period of 6.1d. Variability of this type is well documented for a
variety of stars (e.g. Herbig 1962, Bouvier {\it et al}. 1995, Rydgren \&
Vrba 1983). The puzzling property of a number of the classical TTS (CTTS)
is that their variability period {\it changes} on a time scale of weeks to
months (BP-Tau, mentioned above, is the only case where the period change was
actually {\it continuosly} monitored).

One type of explanations of this effect involves the stellar magnetic field,
following the idea of Ghosh \& Lamb (1978) (see also K\"{o}nigl, 1991). An
oscillation of sorts is assumed to occur in the Alfv\'{e}n radius and thus
the magnetic boundary layer there changes its size and position, causing the
variability (Smith { \it et al. }1995, Armitage 1995). The weakness of these
models lies in the need for an organized strong dipole magnetic field in
TTS, which is rather doubtful observationally (Menard 1996). Another
promising model is that of cool and hot spots on the stellar surface, with
cool spots analogous to sunspots and hot spots being the by-product of the
accretion process (Vrba {\it et al.} 1986, Bouvier \& Bertout 1989, Bouvier
{\it et al}. 1993, 1995).

A recent series of observations, the COYOTES{\footnote{%
Coordinated Observations of Young stellar ObjecTs from Earthbound Sites.}}
II, were obtained over a 2-month monitoring of the light variations of TTS
of the Taurus-Auriga dark cloud. The results provide further evidence for
temporal variations of the photometric periods of CTTS. For example, IQ Tau
and GM Aur exhibit relatively flat lightcurves with narrow minima, while the
other lightcurves are more sinusoidal. DR Tau is a special case, as both the
length and depth of the minima vary, and the lightcurve is neither
sinusoidal nor flattened, but exhibits mixed traits. In the CW Tau case, the
star becomes {bluer} in U-V and B-V when {fainter} than V=$14.0^m$. It also
exhibits large photometric variability, with amplitude variations ranging from
almost $3^{m}$ in the U-band to $1.5^{m}$ in the I-band.

Several photometric periods were reported for DF Tau at different epochs:
8.5d (Bouvier \& Bertout 1989), 7.9d (Richter {\it et al.} 1992), and 9.8d
in the COYOTES II observations. DR Tau and DG Tau may be also candidates for
this category. In the COYOTES I campaign two possible periods were found for
DR Tau (2.8d and 7.3d), while the COYOTES II data points to 7.23d; the DG
Tau data from COYOTES I/II reveals a possible period of 6.3d, while a period
of 4.3d is claimed by Guenther(1994, priv. comm with J.Bouvier). From the
data for TAP 57NW it also seems that the period and the overall shape of the
lightcurve vary with time. In addition to these period variations, the same
stars sometimes do not exhibit {\it any} periods in their lightcurve (e.g. DF Tau
- Rydgren {\it et al}. 1984, Bouvier \& Bertout 1989, BP Tau- Simon {\it et
al}. 1990), indicating that the period not only changes but may disappear
temporarily.

Only stars whose lightcurves are consistent with the presence of hot spots
were found so far to exhibit period changes and disappearances. This suggests
that the cause of these phenomena is to be found in the accretion process
itself. This is also suggested by the spectroscopic signature of mass inflow
at free fall velocities (Hartmann {\it et al}. 1994, Edwards {\it et al.}
1987), which provide support for the magnetic accretion mechanism.

The period variations cannot be attributed to true changes in the stellar
rotation, as the response time of the stellar angular velocity to non-steady
accretion is of the order of $10^5y$ (Pudritz \& Patel 1994) while for
example the timescale over which the period changes are observed to occur is
two weeks for BP Tau (Simon {\it et al.} 1990). Thus one is bound to
conclude that the cause of the period variations lies in changes of the
angular velocity of the hot spots themselves.

\section{The model}

King \& Regev (1994) (hereafter KR) proposed that the low rotation rates and
outflows in TTS could be explained given the presence of a magnetic loop
structure on the central star in a T Tauri-disk system. The loop is assumed
to interact with the disk material via a magnetic surface drag force exerted
on diamagnetic gas blobs, following the scheme of Drell, Folley and Ruderman
(1965). These authors showed that the energy-loss timescale for a
diamagnetic object of mass $m$ crossing the lines of a magnetic field $B$ is
\begin{equation}
t_{drag}=\frac{c_Am}{B^2l^2}\;,
\end{equation}
with $l$ a typical length scale for the object and $c_A$ the Alfv\'{e}n
velocity in the plasma outside the object. The minimum conductivity
condition for this equation to hold is
\begin{equation}
\sigma >\frac{c^2}{2\pi lc_A}\;.
\end{equation}
This reduces to a condition on the blob lengthscale which is satisfied for all
realistic parameters appropraite for TTS conditions (see KR).

Defining now a drag coefficient $k=t_{drag}^{-1}$, the drag force per unit
blob mass is written as
\begin{equation}
{\bf f}_{drag}=-k\;[{\bf {v}^{\prime }}-({\bf {v^{\prime }}}\cdot \hat{{\bf b%
}})\hat{{\bf b}}],
\end{equation}
where ${\bf {v}^{\prime }}$ is the blob velocity relative to the field
lines and $\hat{{\bf b}}$ is a unit vector in the direction of the
field. ${\bf f}_{drag}$ is thus perpendicular to the field lines and
its manitude is proportional to the perpendicular velocity component.
The drag coefficient, $k$,  depends on the magnetic field strength and on the local
interblob plasma density (through $c_A$). Therefore it is a function of
position. In addition, it depends also on the individual blob parameters.

We shall use a magnetic field of a flux tube resulting from a stellar
magnetic loop crossing the disk, with the following functional dependence on
position:
\begin{equation}
B(x,y,z)=B_0 e^{ \left( {-\frac{(x-x_0)^2+(y-y_0)^2}{2w_p^2}}\right)} e
^{\left( -\frac{z^2}{2w_z^2}\right)} \;,
\end{equation}
where $x,y,z$ are Cartesian coordinates (the origin is at the star's center
and $z=0$ is the disk plane), $B_0$ is a
constant and $w_p\;,w_z$ are the Gaussian spreads characterizing the loop.

The functional form of the field should approximate the spatial dependence
of a magnetic field localized in a flux tube with a horizontal spread $w_p$
around the point $x_0,y_0$. A Gaussian form is chosen as a convenient
possibility. The vertical cutoff (Gaussian decline in the field strength,
with spread $w_z$) reflects our wish to avoid repeated blob-flux tube interaction  in our
simple model. In any case, a (Gaussian) vertical cutoff in $k({\bf r})$ will
follow from its dependence on the disk density (see below), which is
usually assumed to have a Gaussian decay with height.

The full expression for the drag coefficient, $k$, is obtained by substituting
$c_A=B/\sqrt{4 \pi \rho}$ and $m=l \rho_b$ in the original definition.
$\rho_b$ and $\rho$ are the densities of the blob and the interblob
plasma in the disk, respectively. Thus
\begin{equation}
k=2\sqrt{\pi} B \rho^{1 \over 2} {1 \over \rho_b l}\;.
\end{equation}

The equations of motion of a blob are written in a frame
rotating with the star, with angular velocity $\vec{\omega}$ as
\begin{equation}  \label{eq:motion}
{\ddot{{\bf r}}}=-\frac{G M {\bf r}}{r^{3}}-
2 {\vec{\omega}} \times \dot{\bf r}
-{\vec {\omega}} \times ({\vec \omega} \times {\bf r})
- k \left( \dot{\bf r}
-(\dot {\bf r }\cdot{\hat{{\bf b}}}){\hat{{\bf b}}} \right) \;,
\end{equation}
with ${\bf r}\equiv (x,y,z)$
and ${\bf \hat b}=(b_{x},b_{y},b_{z})$ defined
as
\begin{equation}
b_{x}=\sin \theta \cos \phi \;\;, b_{y}=\sin \theta \sin \phi \;\;, b_{z}=\cos
\theta \;.
\end{equation}
$\theta $ and $\phi $ are angular parameters corresponding to the flux tube
direction relative to the disk. $M$ is the cental star's mass and $G$
the gravitational constant.

We first assume that ${\vec {\omega}} = \omega {\bf{\hat z}}$, i.e. the
rotation axis of the star is the $z$-axis of the coordinate frame
($\hat {\bf z}$ is the corresponding unit vector). This is very
reasonable and means that the star and disk rotate around the same axis.
Next we
nondimensionalize the equation of motion, scaling the physical
variables by the following chracteristic units:  distances are scaled by
the value of the corotation radius, 
defined here to be the radius at which the local Keplerian angular velocity
equals $\omega $-the angular velocity of the star and thus the field,
$r_{co} = (GM/\omega^2)^{1/3}$; the
time by the rotational time $\tau=1/\omega$ and the rotational velocity
by $\omega$. The following nondimensional equation of motion then results:

\begin{equation}  \label{eq:motion2}
{\ddot{{\bf r}}}=-\frac{{\bf r}}{r^{3}}-
2 {\bf{\hat z}} \times \dot{\bf r}
-{\bf{\hat z}} \times ({\bf {\hat z}} \times {\bf r})
- \alpha \kappa({\bf r}) \left( \dot{\bf r}
-(\dot {\bf r }\cdot{\hat{{\bf b}}}){\hat{{\bf b}}} \right) \;,
\end{equation}
where the nondimensional drag coefficient is
\begin{equation}
\kappa({\bf r})=\frac{B({\bf r})}{B_0} \sqrt{\frac{\rho(z)}{\rho_0}}\;.
\end{equation}
Here we have assumed that the interblob disk density depends on $z$
only and $\rho_0$ is the midplane value. As mentioned before this
causes the appropriate decline of $\kappa$ with height.
The constant parameter $\alpha $ is
\begin{equation}
\alpha= \pi^{-1/2} P_{\star} B_0 \rho_0^{1/2}\frac{1}{\rho_b l}\;,
\end{equation}
where $P_\star=2\pi/\omega$ is the rotation period.
With the various physical variables expressed in their typical values
the parameter $\alpha$ can be written as
\begin{eqnarray}
\alpha =0.4875 \frac{B_0}{100G}\left( \frac{\rho _0}{10^{-10}[g/cm^3]}\right)^{1/2}
P_{\star }[days] \nonumber \\ 
\times\left(\frac{l}{10^9[cm]}\frac{\rho _b}
{10^{-7}[g/cm^3]}\right) ^{-1}.
\end{eqnarray}

The interaction is thus parametrized by the interaction strength $\alpha $,
the flux tube width $w_p$ and the tube direction $\theta ,\phi $. $w_z$
should be approximately the disk half-thickness. The motion of a blob after
suffering an interaction with a magnetic loop will obviously depend also on
the initial blob velocity relative to the field line. As a rule the blobs
will be taken to orbit the central star with a local Keplerian velocity plus
a small inward drift, hence the importance of the interaction radius $r_0=
\sqrt{x_0^2+y_0^2}.$ 

Eq. (\ref{eq:motion2}) describes the blob motion in a rotating frame, with
the Coriolis and centrifugal terms included as appropriate. It can be solved
numerically for a variety of initial conditions and loop and blob
parameters. We shall refer to the solutions of this equation as ballistic
orbits if the trajectory leaves the disk and crosses its plane again.

\setcounter{equation}{0}

\section{Ballistic orbit calculations}

\subsection{General considerations}

We start the integration of the blob motion from a point close to the
interaction point $(x_0,y_0)$ in the disk plane, but far enough from it so
that the magnetic drag term is negligible. A distance of several $w_p$ is
appropriate and thus with $w_p=\epsilon r_0$ and $\epsilon \ll 1$, we obtain
a sharply (with respect to $r_0$) increasing field.

For $r_0<r_{co}$ the blobs are slowed down by the magnetic drag (see KR). If
no vertical velocity component is acquired this will merely increase the
accretion rate. However, significant vertical velocity component may often
be produced by the interaction; in such cases the blob leaves the disk -
only to fall back on it (close to the star in most cases), after a very
inclined orbit.

In our calculations the blob is started with a Keplerian velocity (in the
inertial frame) plus a small inward drift. For the case $r_0<r_{co}$ the
blob is started behind of flux tube, so as to let the blob overtake the
tube. In this work we shall concentrate only on this case, since $r_0>r_{co}$
implies almost always ejection of the blob out of the disk plane
without subsequent impact, see KR and Pearson \& King (1996). As
pointed out in the Introduction, we focus here on the possibility of blob
ejection out of the disk plane, but with less than escape energy and with
the blob supposed to land quite close to the star (see below). Indeed we
were able to obtain such a situation, for a large fraction of the parameter
space, in the case $r_0<r_{co}$.

For a given choice of the parameters, each ejected blob will follow exactly
the same path and thus its center of mass will impact the disk at the same
position (in the rotating frame) as all the preceding ones. If we envisage a
succession of blobs being ejected one after the other and forming during
their flight a continuous stream of matter, this stream will impact the disk
at a fixed (in the rotating frame) position. The impact point with the
aforementioned hot spot created by the stream will thus move on the disk,
around the star, with an angular velocity $\omega $. If the spot is close
enough to the star for a given inclination angle, it will be, at least
partly, eclipsed by the star. The observed luminosity variability due to
this spot will then have the fixed frequency $\omega $. The stream may also
impact the star and the observational result, regarding time variability,
should be the same. Now, if we allow the field parameters to vary in time,
the spot is bound to move, in the rotating frame, on a timescale of the
order of the magnetic field variation timescale. We shall assume the this
timescale approximately of the order $t_B\sim 5P_{\ast}$, where $P_{\ast}
=2\pi /\omega $ is a rotation period of the field. Thus we postulate that
the field configuration changes (for example due to the interaction with the
disk plasma and the blobs, or due to internal changes), and this happens on
a time scale of $\sim 5$ rotations. A spot moving in the rotating frame can
explain all the unusual features in the observed light curves. The period
can change on this time scale. The variability may disappear (if the spot
moves far enough out as not be eclipsed any more) and reappear again, if the
spot moves back in. A systematic study of all the possibilities calls for a
many body particle numerical simulation for the blobs in the disk. In
addition we need to simulate numerically the stream of the ejected matter
and its impact on the disk. Such a project is now in progress, while here we
would like only to demonstrate the idea by describing various cases of
individual blobs.

If we fix all the parameters but start the calculation with different
initial conditions (due to blobs coming from slightly different places), we
expect a spread in the impact point in the rotating frame, giving rise to an
extended hot-spot. The typical spread of impact points we have found is of
the order of the stellar radius, $R_{\star }$ for the cases described below,
with the initial blob conditions around the magnetic flux tube set to give a
ballistic orbit.

\subsection{Results for varying field parameters}

The canonical values of the parameters for our calculations are: $\theta
=\pi /4$, $\phi =3\pi /4$, $r_0=0.85$, $\alpha =100$ in non-dimensional
units, with time scaled by $P_{\star }$ and lengths by the value of the
co-rotation radius $r_{co}$. We chose always a single blob initial
condition, remembering that a spread in the spot size is expected due to
blobs with close initial conditions. We then vary (linearly in time) each of
these parameters in turn, keeping the others fixed. The timescale of change
is of the order of $5P_{\star }$. The Gaussian spreads are always kept fixed
at $w_p=w_z=0.05$.

The variation of $\alpha $ has a relatively small effect on the impact point
position in the rotating frame. Thus the spread in blob sizes and masses as
well as the actual absolute value of the field matter little in this model
(A variation of $\alpha $ between $1$ and $10^6$ resulted in a change of
only $\sim 2\%$ in spot angular velocity, see figure \ref{fig:omega}(a).

Significant change occured when we have varied the value of $r_0$, the radius
of the point where the loop crosses the disk (in units of $r_{co}$). A
linear change $r_0=0.6+0.08t$ (variation in the range $0.6-0.92$ during $4$
rotations) resulted in changes of the spot center position in polar
coordinates ($R_{imp},\Theta _{imp}$), shown as a function of time in figure
\ref{fig:coord}(b). Unlike in the variable $\alpha $\ case, (figure \ref
{fig:coord}(a)), here the apparent angular velocity of the spot changed
drastically; combined with the spot grazing the star itself part of the time
($R_{\star }\simeq 0.23r_{co}$) this produced a lightcurve very much
resembling that of DR Tau (see fig. 3 of Bouvier {\it et al}. 1995). The
model lightcurve exhibits variation in shape, period and depth of the minima
(figures \ref{fig:spectrum},\ref{fig:r0curve})

By varying $\phi $ and $\theta $ we again obtain significant motion of the
spot in the rotating frame. Figure \ref{fig:coord} depicts these cases as
well: spot's coordinates evolution with $\theta $ varying as $\theta
=0.5+0.2t$- curve (c), and the coordinate evolution with $\phi =1.9+0.3t$-
curve (d). We can see that a spot moves away from the star, and so the
possibility of period disappearance and its consequential return is present,
although with a different apparent angular frequency. It should be also
noted that although $\Omega$ (spot's angular velocity as seen by an observer
in an inertial frame) varies some $10-15\%,$
the angular velocity deduced from the spectrogram is just the{\ average.} In
both cases the apparent period differed by about $10\%$ from $P_{\star }$%
.(fig. \ref{fig:spectrum})

\subsection{Model predictions}

In order to be able to predict whether the variability can be observed, one
needs to know the stellar radius, $R_{\star }$, the angle of inclination of
the disk and the spot size. The apparent angular velocity of the spot is
clearly $\Omega =\omega +\dot{\Theta}_{imp}$. This can be calculated with no
further assumptions. We assume that the spot size is very small ($\sim
0.2R_{\star})$ and the inclination angle is such that eclipses are {always}
occurring for the cases studied. The star is assumed to have $R_{\star
}=2R_{\odot }$, $\Omega _{\star }=\omega =0.1\Omega _{K_{\star }}$ (a tenth
of the breakup value). The results should thus be considered as illustrative
only. In figures \ref{fig:omega},\ref{fig:spectrum} we see the angular
velocity of the spot in the inertial frame, i.e. as should be observed, the
and the power spectrum of the predicted intensity variation-- for the three
cases of varying different parameters (corresponding to the results depicted
in figure \ref{fig:coord}).

\section{Discussion}

The above model is an extremely simplified view of the whole problem. In
reality, a process for the creation of the diamagnetic blobs would have to
be specified; the interaction between blobs should be included;
and the disk thickness has to be taken into account, as it is unreasonable
to expect that a blob flung out from the downward side upwards will traverse
the disk without colliding with the disk and other blobs inside it. Here we
do not specify the MHD instability responsible for the creation of blobs,
but only note that in magnetic Cataclysmic Variables there is probably
observational evidence for a blobby accreted plasma (see Wynn and King
1995). In addition, we need the blobs to exist only for a time sufficient
for the magnetic interaction in the disk.

If the conditions confining the blobs are specific to the disk, they will
expand on the thermal time-scale after ejection. Expanding, they will merge
together into a stream-like structure; thus the above picture of single
blobs falling into the disk is useful only in the context of checking the
whole idea; the results should be taken as no more than qualitative and
representative of the general scales.

Still, the model looks promising, as it might explain the variability of the
observed period in certain TTS; for example, the disappearance of the period
altogether can be explained by the hot spot moving farther away from the
star, thus evading being eclipsed. The model works only for highly inclined
systems. It will be thus interesting to look for an observational
correlation between period variability and inclination in CTTS. We do not
rule out stellar spots as causing variability in the case when disks are
absent (then this is the only effect) or when they are present (then the
star and disk spots are both present, but only the disk spots are
responsible for period changes).

In the next step of this research we intend to perform a three dimensional
SPH simulation of the accretion disk and the stream resulting from the
magnetic interaction taking into account in the mass infall rates as well
(which shall affect the general spot luminosity).

Concluding, we make the following remarks:

\begin{enumerate}
\item  The optical thickness of the stream and the penetration depth in the
disk are functions of the stream density. We expect that if the stream
density gets smaller, the observed temperature will get closer to the shock
temperature; meanwhile the total luminosity will decline. This could explain the
observed blue shift during the low luminosity phase in some TTS.

\item  In some instances the observed period disappears altogether, and
reemerges afterwards. In our model, for low disk inclination angles $i$, the
spot will not be eclipsed for $R_{imp}>R_{\star }/\sin (i)$. As we see in
the examples, $R_{imp}$ varies in a wide range, so this is quite plausible.
Another possible explanation is that the stream disappears altogether (for
example if the loop lies in the disk plane), and reappears on the other side
of the disk, its luminosity greatly diminished by the disk thickness. In
addition, although improbable, the spot angular velocity could be very close
to $0$ if a suitable parameter variation occurs. The spot could thus be
visible for a long duration.

\item  We do not take here into account the scenario where the spot falls on
the star itself. This occurrence is quite possible, especially when taking
into account the spot size. At least a part of the spot may be on the star,
while the other part is still on the disk. For instance, a disk spot
changing into a star spot will have the following influences on the
lightcurve: as a growing proportion of the stream falls on the star, the
apparent area of the spot will grow; thus the luminosity will rise. Being on
the stellar surface, the spot apparent size will vary as $\sin (\Omega
_{spot}t)$, and the luminosity will vary sinusoidally. We may expect that
the lightcurve will look somewhat similar to the one we obtained in figure
\ref{fig:r0curve}.

\item  If the spot grows beyond $2R_{\star }$, we shall encounter shallower
minima as only a part of the spot will be eclipsed. Taking into account the
fact that a star is spherical, even for a constant spot size the eclipsed
part shall decline as the spot falls further from the star, again resulting
in shallower (and shorter) eclipses.
\end{enumerate}

\acknowledgments

This work was supported by a grant from the Israel Science Foundation of the
Israel Academy of Sciences and by the Minerva Center for the Physics of
Complex Systems. O.R. acknowledges the hospitality of J. Fourier University
at Grenoble and C.B. the hospitality of the Institute of Theoretical Physics
at the Technion. \\\\

\newpage

\figcaption[coord.eps]{Spot coordinates (in the rotating frame) as a function of time (a)
for varying $\alpha $ ($\alpha $ in the range $1-10^6$), (b) $r_0$(range
0.6-0.92 $r_{co}$),(c) $\phi $ (range 1.9-3.1 rad),(d) $\theta $ (in the
range 0.5-1.3 rad). t -- time in $P_{\star }$ units
\label{fig:coord}}

\figcaption[omega.eps]{Apparent spot angular velocity, with the same parameters as figure
1. We can see that the maximal variation occurs in the $r_0$ case (b), while
varying $\alpha $ hardly infuences the results (a). Variation of $\phi $ (c)
and $\theta $ (d) produce a comparable variability of some 10-15\%. (t --
time in $P_{\star }$ units)
\label{fig:omega}}

\figcaption[spectrum.eps]{Power spectrum of the calculated intensity variations. (a)In the
varying $\alpha $ case the frequency obtained equals 0.9 $2\pi /\Omega
_{\star }$ (b) Varying $R_0$ case. The deduced frequency is half the actual
one, as one of the minima would be inobservable. (c) Varying $\phi $ case.
The power spectrum suggests a frequency that is 10 \% less than the actual.
(d) Varying $\theta $ case. The deduced frequency is about 15\% more than
the actual.(t -- time in $P_{\star }$ units)
\label{fig:spectrum}}

\figcaption[r0curve.eps]{Modelled spot luminosity (variable $r_0$ case). Both the period and
the eclipse length vary, as does the overall shape of the light curve. The
second minimum (arrow) is too shallow to influence the power spectrum.
\label{fig:r0curve}}


\addcontentsline{toc}{chapter}{\bf Bibliography}

\begin{thebibliography}{99}
\bibitem{armitage}  Armitage K.J.,1995, \mnras,  {\bf 274},1242

\bibitem{}  Bertout C., 1989,\araa, {\bf 27},351

\bibitem{}  Bouvier J., Bertout C., 1989, A\&A,{\bf 211},99

\bibitem{}  Bouvier J., Cabrit C., Fernandez M., Martin E.L.,Matthews
J.,1993, A\&A, {\bf 272}, 176

\bibitem{COYOTES}  Bouvier J., Covino E., Kovo O., Martin E.L.,Matthews
J.M., Terranegra L., Beck S.G., 1995, A\&A, {\bf 299 }, 89

\bibitem{}  Drell, S.D., Foley H.M., Ruderman, M.A., 1965, J. Geophys. Res.,%
{\bf 70}, 3131

\bibitem{}  Edwards S., Cabrit S., Strom S.E., Heyer I., Strom K.M.,
Anderson E. 1987, \apj, {\bf 321}, 473

\bibitem{}  Ghosh P., Lamb F.K., 1978, \apj, {\bf 223}, L83

\bibitem{}  Herbig, G.H., 1962, Adv. in A\&A.,{\bf 1},47

\bibitem{}  Hartmann L., Hewett R., Calvet N. 1994, \apj, {\bf 426}, 669

\bibitem{KR}  King, A.R., Regev, O., \mnras, {\bf 268}, L69-L73 (1994)

\bibitem{}  K\"{o}nigl A., 1991, \apj, {\bf 370}, L39

\bibitem{}  Lin, D.N.C., Pringle, J.E., 1976, IAU Symp. No.7, p183, eds.
Eggleton,P., et al., Reidel, Dordrecht, Holland.

\bibitem{}  Menard, F., 1996, Magnetic Effects in Accretion Workshop,
Technion, Haifa.

\bibitem{KP}  Pearson,K.J., King, A.R. , 1995, \mnras, {\bf 276}, 1303

\bibitem{}  Pudritz R.E., Patel K.. 1994, \apj, {\bf 424} 688

\bibitem{}  Richter M., Basri G., Perlmutter S., Pennypacker C. 1992, \pasp, {\bf 104}, 1144

\bibitem{}  Rydgren A.E.,Vrba F.J., 1983, \apj, {\bf 267} 191

\bibitem{}  Rydgren A.E.,Zak D.S.,Vrba F.J.,Chugainov P.F,Zajtseva G.V.,
1984, \aj, {\bf 89} 1015

\bibitem{}  Simon T., Vrba F.J., Herbst W. 1990, \aj, {\bf 100(6)},1957

\bibitem{smithbonnell}  Smith, K.W., Lewis, G.F., Bonnell, I.A., \mnras, {\bf %
276}, L5-L8 (1995)

\bibitem{}  Vrba F.J., Rydgren A.E {.,} Chugainov P.F., Shakovskaya N.I.,
Zak D.S., 1986, \apj,{\bf 306}, 199

\bibitem{}  Worden S.P., Schneeberger T.J., Kuhn J.R, Africano J.L., 1981,
\apj, {\bf 244}, 250

\bibitem{kingwynn}  Wynn G.A. and King A.R., 1995,\mnras, {\bf 275},9
\end{thebibliography}
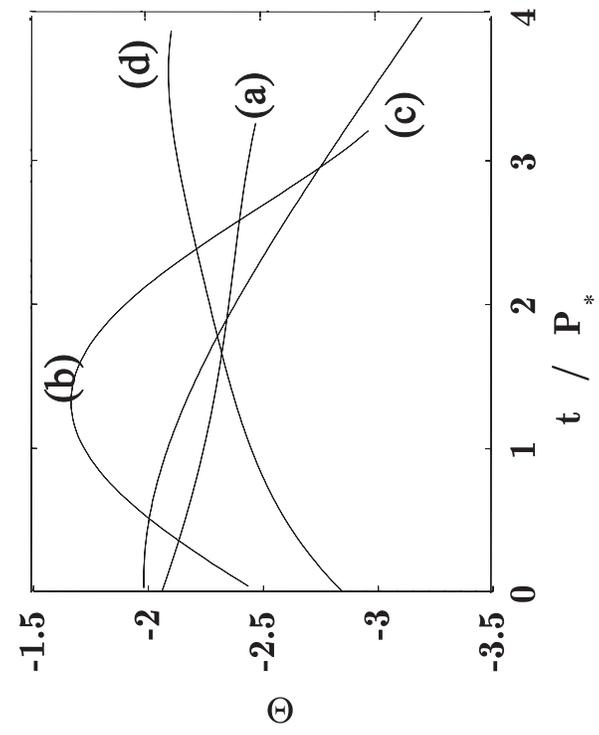
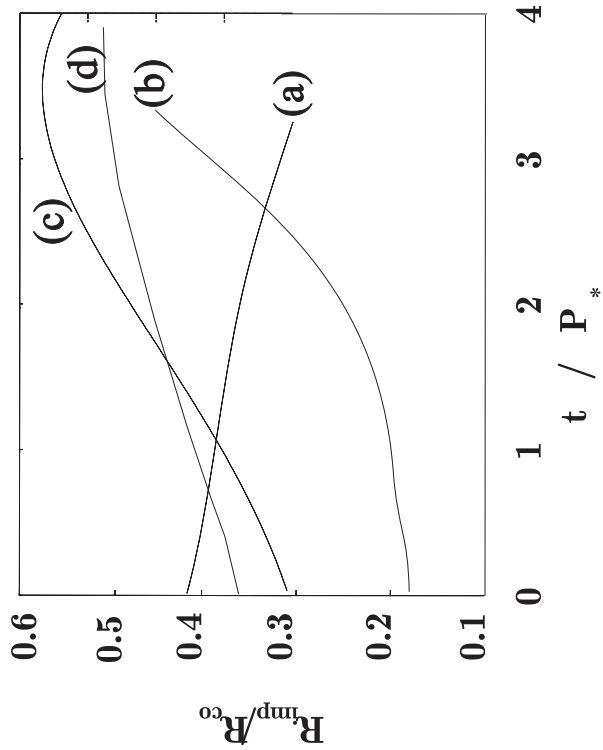
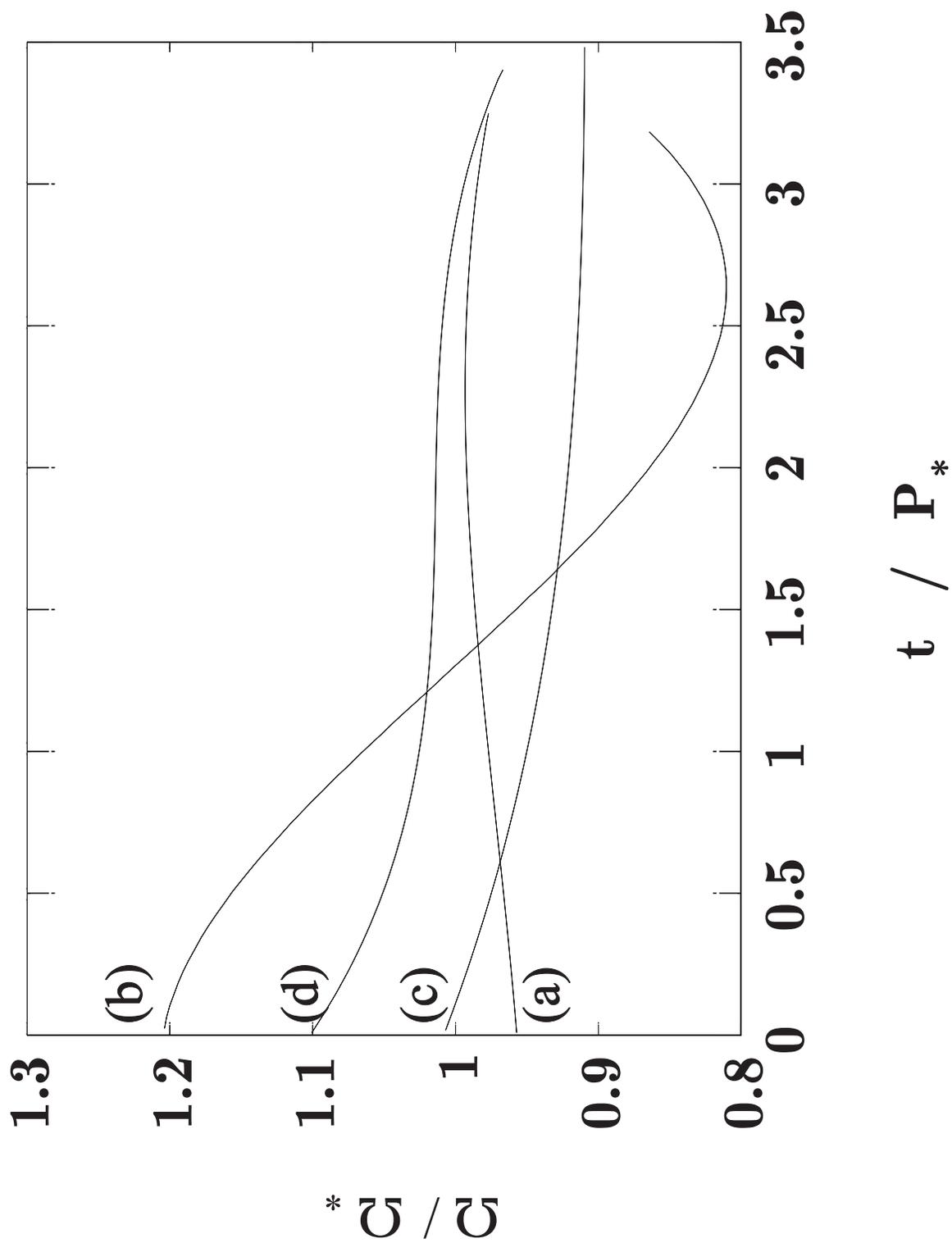
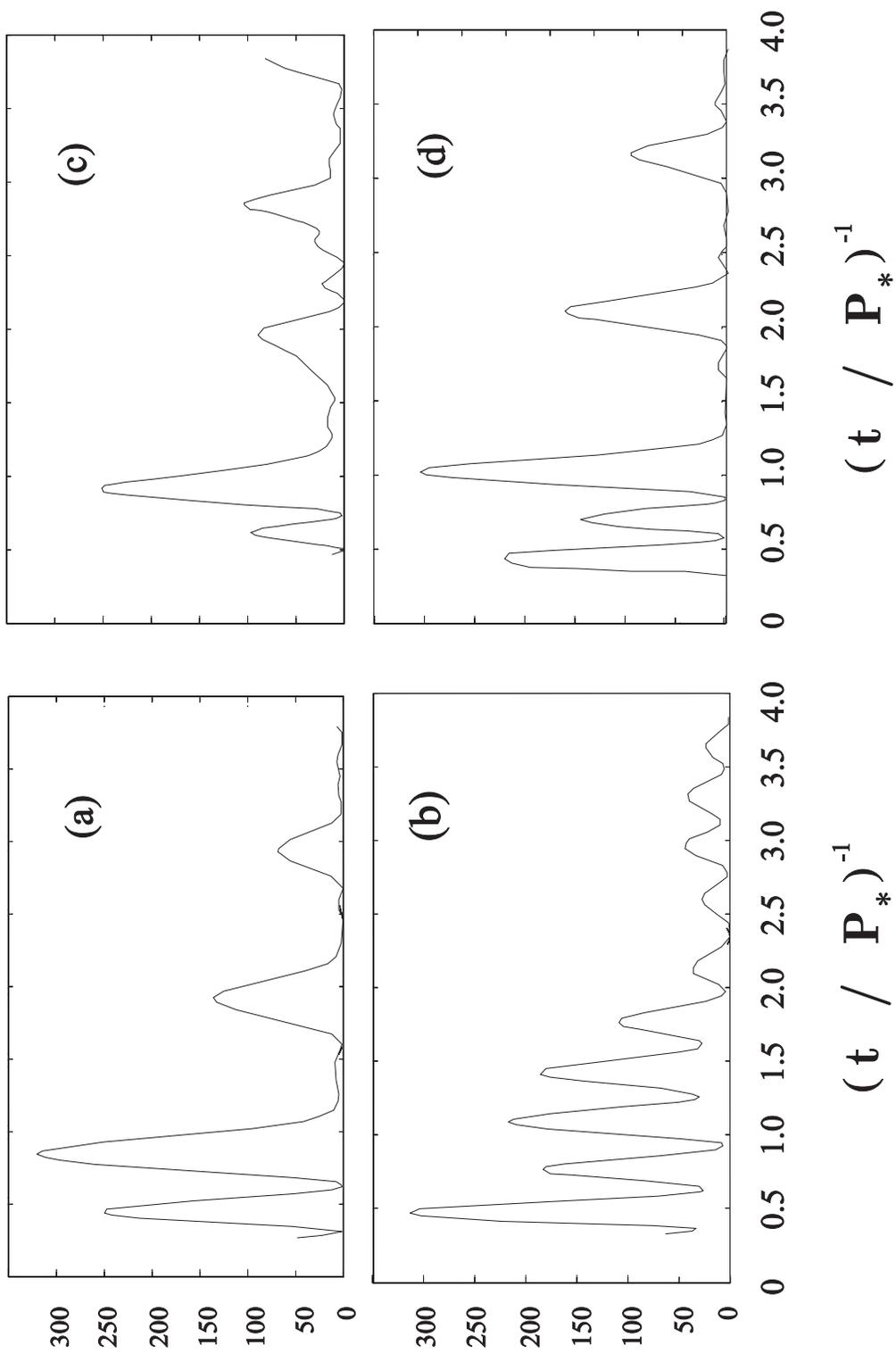
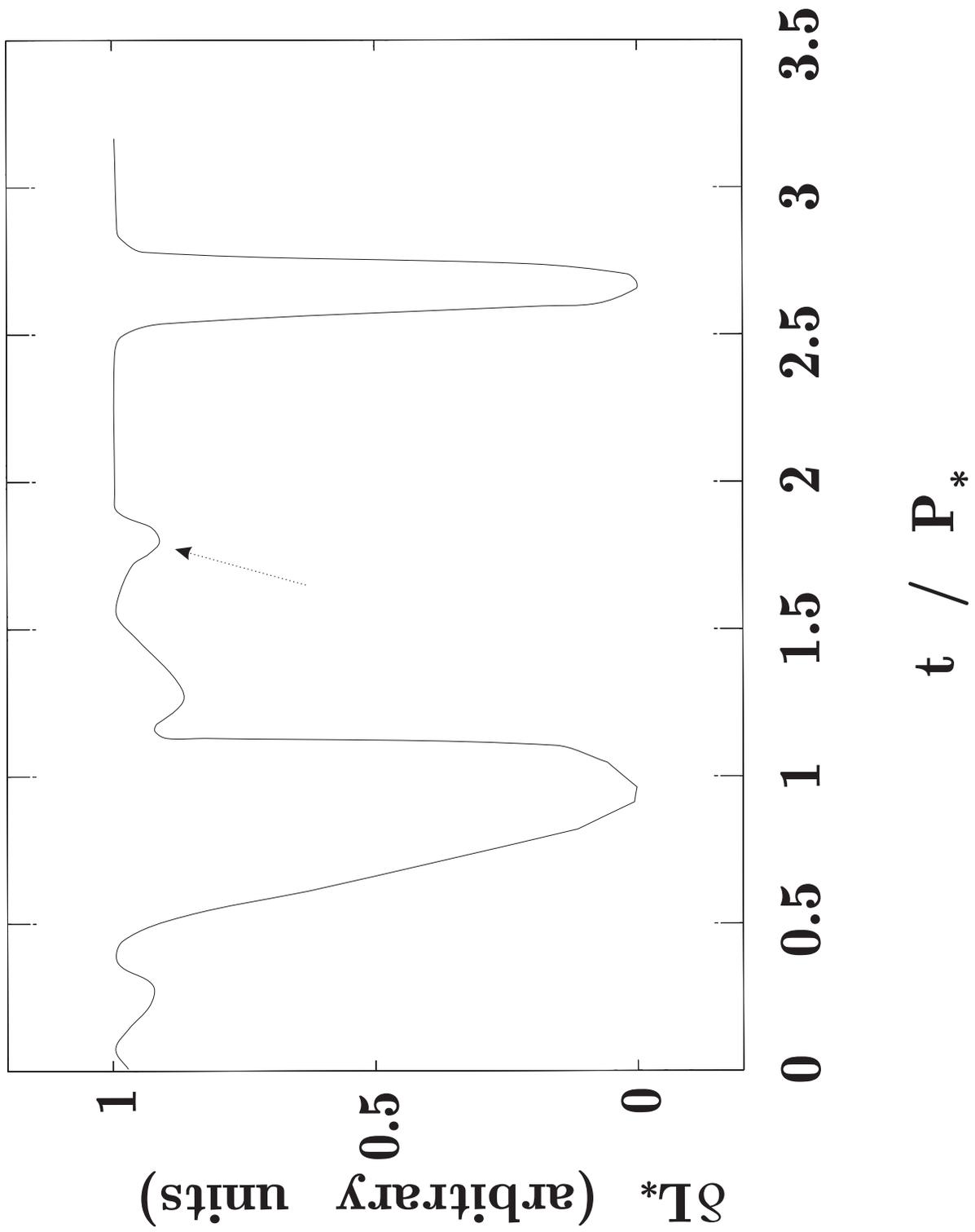
\end{document}